\documentclass[12pt]{article}

\usepackage{graphicx}
\usepackage{color}
\usepackage[colorlinks=true, urlcolor=blue, linkcolor=blue, 
citecolor=blue]{hyperref}
\usepackage{cite}

\newcommand{\beq}{\begin{equation}}
\newcommand{\eeq}{\end{equation}}
\newcommand{\beqa}{\begin{eqnarray}}
\newcommand{\eeqa}{\end{eqnarray}}
\newcommand{\beqar}{\begin{eqnarray*}}
\newcommand{\eeqar}{\end{eqnarray*}}

\begin{tiny}

\end{tiny} 

\begin{document}
\thispagestyle{empty}
$\,$

\vspace{32pt}

\begin{center}

\textbf{\Large The Sun at TeV energies:\\
gammas, neutrons, neutrinos and a cosmic ray shadow}

\vspace{50pt}
Miguel Guti\'errez, Manuel Masip 
\vspace{16pt}

\textit{CAFPE and Departamento de F{\'\i}sica Te\'orica y del Cosmos}\\
\textit{Universidad de Granada, E-18071 Granada, Spain}\\
\vspace{16pt}

\texttt{mgg,masip@ugr.es}

\end{center}

\vspace{30pt}

\date{\today}

\begin{abstract}

High energy cosmic rays reach the surface of the Sun
and start showers with thousands of secondary particles. 
Most of them will be absorbed by the Sun, but a fraction of the neutral
ones will escape and reach the Earth. 
Here we incorporate a new ingredient that is essential 
to understand the flux of these solar particles: the cosmic ray shadow
of the Sun. We use Liouville's theorem to argue that the only
effect of the solar magnetic field on 
the isotropic cosmic ray flux is to interrupt some of the trajectories that
were aiming to the Earth
and create a shadow. This shadow
reveals the average solar depth crossed by
cosmic rays of a given rigidity. The absorbed cosmic ray flux is 
then processed in the thin 
Solar surface and, assuming that the emission of neutral particles 
by low-energy charged particles is isotropic, we obtain {\it (i)}
a flux of gammas that
is consistent with Fermi-LAT observations, {\it (ii)} a flux of 100--300
neutrons/(year m$^2$) produced 
basically in the spallation of primary He nuclei, and {\it (iii)} a neutrino flux that is 
above the atmospheric neutrino background at energies above 0.1--0.6 TeV 
(depending on the solar phase and the zenith inclination). More precise measurements of the
cosmic ray shadow and of the solar gamma flux, together with the possible 
discovery of the neutron and neutrino signals, would provide valuable
information about the magnetic field, the cycle, and the interior of the Sun.

\end{abstract}

\vfill
\eject

\noindent High energy astroparticles provide a picture of the
sky that complements the one obtained with light at different frequencies. Their study
during more than 100 years has helped us to understand the environment where 
these particles are produced (supernovas, 
pulsars, active galactic nuclei or gamma ray bursts)
and the medium that they find on their way to the Earth. Indeed, we expect 
them from any event where nature reaches extreme conditions,
also the neutron star mergers recently discovered through gravitational waves 
\cite{GBM:2017lvd}. We may distinguish three types of astroparticles: 
cosmic rays (CRs, including atomic nuclei and electrons), gamma rays and neutrinos.
CRs are charged particles, they are the only ones to be 
accelerated in astrophysical processes but lose directionality as they propagate
along the magnetic fields present in the cosmos at all scales. In contrast, 
gammas and neutrinos appear as secondary particles produced when CRs 
interact with matter or light, and they always point to the source. It is then apparent
that the observations in these three channels provide 
different pieces of the same puzzle. 

Here we will focus on an astrophysical object that seems difficult to overlook but that
still keeps some very significant secrets: the Sun. With a temperature around
1 keV, one may think that any solar particles with energy in the 
GeV--TeV range may be a sign of new physics, {\it e.g.}, 
the annihilation of 
massive dark matter particles captured by the Sun. This, however, 
is not necessarily the case. In 1991 Seckel, Stanev and Gaisser \cite{Seckel:1991ffa}
described a flux (SSG flux)
of high energy particles produced by CRs showering in the surface of the Sun.
More recently the Fermi-LAT observatory has 
detected  solar gammas both during a minimum and near a peak
of solar activity (2014--2017) \cite{Abdo:2011xn} (see also \cite{Linden:2018exo}).
The flux exhibits a hard spectrum ($\propto E^{-2}$) between
1 and 200 GeV, with a possible dip at $E\approx 40$ GeV \cite{Tang:2018wqp}, 
and then it seems to drop. 
The flux is 10 times above the diffuse gamma ray background and around
7 times larger than the SSG estimate, although its modulation with the solar cycle
seems a clear indication of its CR origin.

\begin{figure}[!t]
\begin{minipage}{0.45\linewidth}
\includegraphics[scale=0.50]{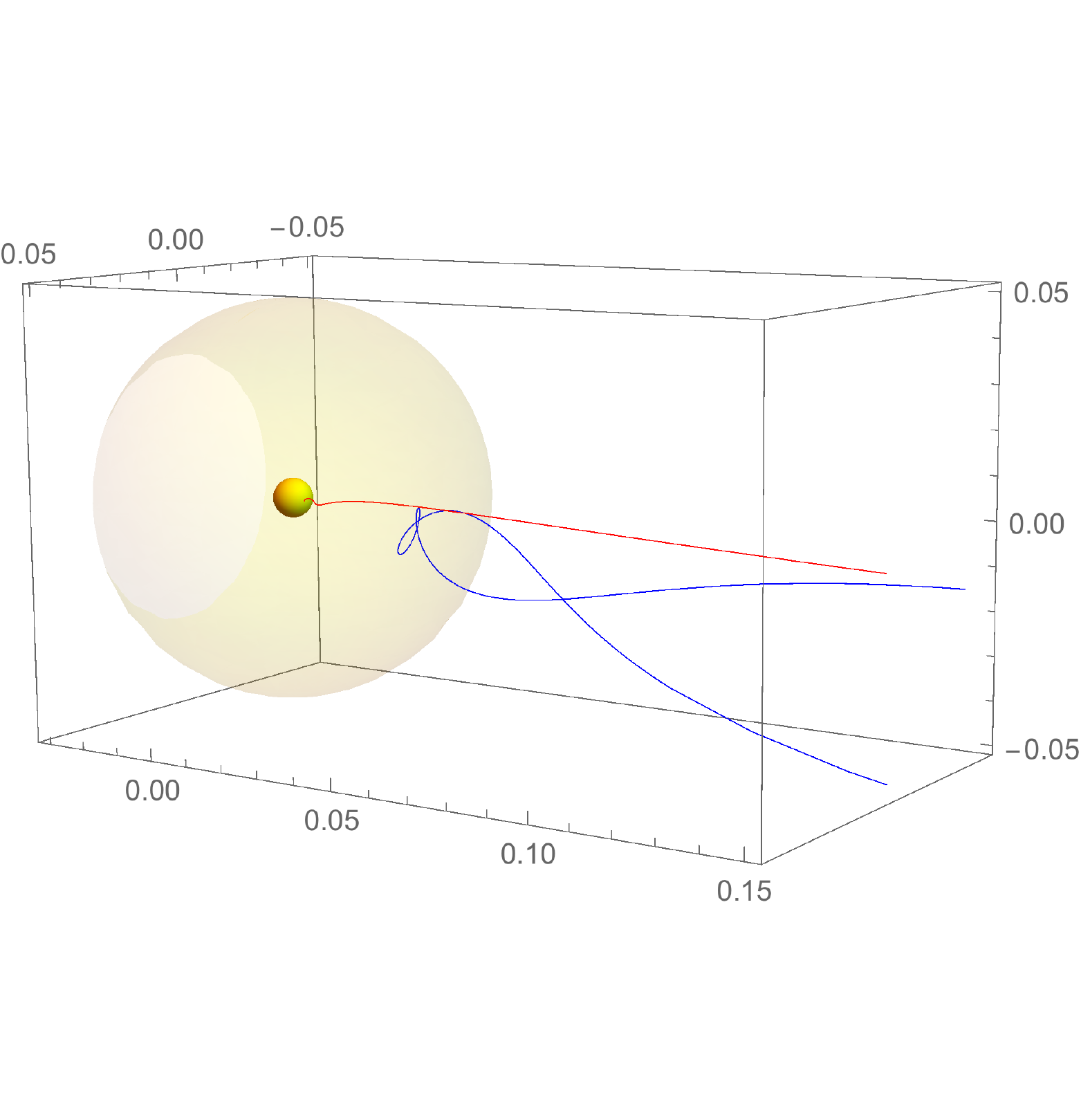}
\end{minipage}
\hspace{1.cm}
\begin{minipage}{0.45\linewidth}
\includegraphics[scale=0.50]{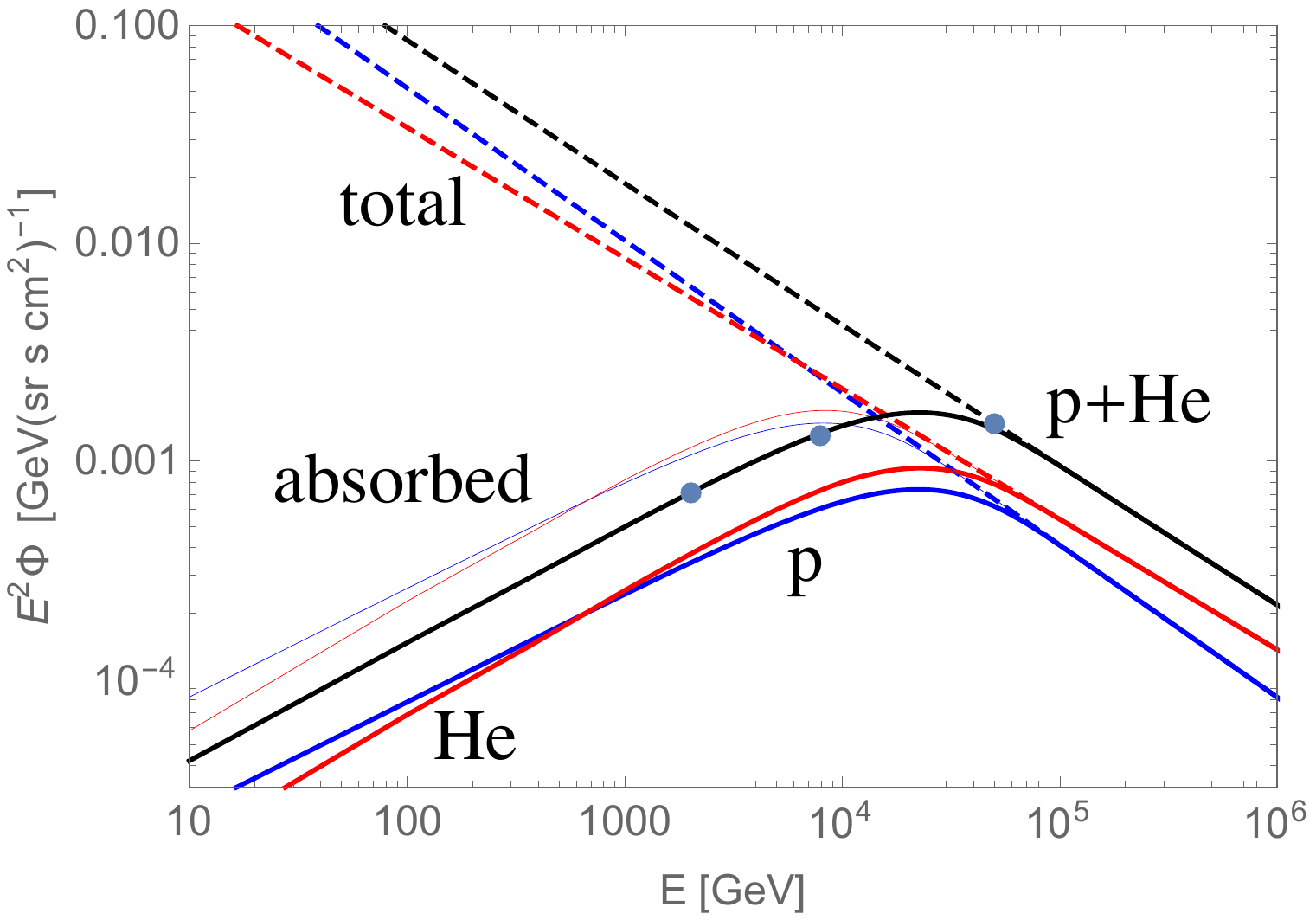}
\end{minipage}
\caption{{\bf Left.} Trajectories through the Parker field 
in the vicinity of the Sun starting at the Earth
for a 5 TeV proton (in AU; closed field
lines starting and finishing on the Sun's surface not included). Only trajectories aligned 
with the open field lines reach distances $R<10 R_\odot$ (shaded
region). 
{\bf Right.} CR flux observed at the Earth
(dashes) and absorbed flux (solid) consistent with HAWC observations (dots). 
Blue (red) lines correspond
to protons (He nuclei) and black lines to p+He.
We include in the plot the absorbed flux expected during a solar minimum (thin solid),
which is obtained by reducing $E_{\rm crit}$ by a factor of 1/3.
\label{f1}}
\end{figure}

The main uncertainty in the SSG calculation was caused by the effect of the solar magnetic
field on the trajectory of CRs as they approach the surface. 
At distances beyond 10 solar radii TeV CRs 
follow ballistic trajectories through the Parker interplanetary field \cite{Tautz:2010vk}, 
which includes 
a strong radial component $\propto R^{-2}$ (open lines) carried by the solar wind. 
Many of the CRs moving towards the Sun 
will experience a magnetic mirror effect before they reach the surface (see Fig.~\ref{f1}).
At closer distances the field lines tend to co-rotate with the Sun ($T\approx 24$ days),
turbulence increases 
and there appears
a new type of (closed) field lines that start and finish on the surface. In addition, this 
magnetism is not stable, it has an 11-year cycle that is correlated with the solar activity. 
Any numerical or analytical calculation of the absorption rate of CRs by the Sun 
seems then uncertain. Instead, we can add a key observation that was not 
available at the time of the SSG analysis: the energy dependent 
shadow of the Sun measured by HAWC \cite{Enriquez:2015nva}. The shadow, with data 
taken during a solar maximum (years 2013--2014), is not a black disk of
$0.27^\circ$ radius (the angular size of the Sun). It is a deficit that appears in the CR flux
at 2 TeV and extends into an angular region ten times larger than the Sun. 
HAWC fits this deficit with
\beq
d(\theta)= -A \,\exp\left(-{\theta^2 \over 2 \sigma^2}\right)\,,
\eeq
providing the parameters $A$ and $\sigma$ at 3 different energies. Integrating 
we obtain that at $2$  and $8$ TeV the deficits are equivalent to 6\% and 27\% of a black disk, 
respectively, whereas at CR energies around $50$ TeV it adds to a 
complete shadow (i.e., a deficit of $\pi \,\theta_\odot^2$, with $\theta_\odot=0.27^\circ$).

To interpret HAWC's results we can use Liouville's theorem, stating that 
the density of CR trajectories along a trajectory in phase space remains constant. 
If we consider a given CR energy and a static
magnetic field (time variations may change that energy), 
then the theorem implies that an isotropic CR flux will 
stay isotropic. A magnetic lens, including a
mirror, will not produce anisotropies, and the only possible effect of the Sun will be to
interrupt some of the trajectories that were aiming to the Earth and create a
shadow. As we turn around the Sun we face different magnetic configurations; the
CR shadow at a given energy 
may appear and disappear, and its apparent position may
change due to the magnetic deflection. At HAWC we see the shadow 
averaged during the observation period and smeared out 
(both by the magnetic deflection and by the experimental error) into an angular region 
of $2.6^\circ$ radius.

The integrated deficit in the CR shadow at different energies reveals then the fraction of
the CR flux absorbed by the Sun. We can model this absorbed flux by assuming that 
the proton trajectories of energy $E$ closest to the Sun 
and aiming to the Earth through the solar magnetic 
field cross an average depth (column density of solar matter) $\Delta X_{\rm H}$, with
\beq
\Delta X_{\rm H}(E)= c_{\rm H} \left({E\over 1\;{\rm GeV}}\right)^{1.15}\,,
\eeq
and $c_H=4.4\times 10^{-4}$ g/cm$^2$. 
It follows that Helium nuclei of energy $2E$ ({\it i.e.}, with 
the same magnetic rigidity) will cross exactly the same average depth, 
$\Delta X_{\rm He}(E)=\Delta X_{\rm H}(E/2)$. 
The 
energy dependent probability of absorption by the Sun $p^{\rm H}_{\rm abs}$ is then
\beq
p^{\rm H}_{\rm abs}=1-\exp{\left(-{\Delta X_{\rm H}\over \lambda_{\rm int}^{\rm H}}\right)}\,,
\eeq
where $\lambda_{\rm int}^{\rm H}$ is the inelastic interaction length (in g/cm$^2$)
for a proton in the Sun. An 
analogous expression would describe the absorption probability of  He nuclei.
In Fig.~\ref{f1} (right) we plot the absorbed flux that we obtain under this hypothesis, 
which is consistent with 
HAWC's observations. We have taken a primary CR flux dominated by
protons and He nuclei with slightly different spectral index ($-2.7$ and $-2.6$, respectively),
neglecting heavier nuclei. At high energies the
total and the absorbed CR fluxes coincide (the CR shadow is complete), whereas
at lower energies an increasing fraction of CRs is unable to
reach the Sun's surface (the shadow disappears). Notice that 
He nuclei have charge $+2$ and will find it more difficult than
protons to reach that surface, but their inelastic interaction length is shorter and the
two effects tend to compensate: absorption becomes important at a similar energy
for both components.

Once absorbed the CR flux will be processed 
into secondary particles. One important observation
is that the environment where the shower develops is very thin, much thinner than the
Earth's atmosphere:
from a vertical trajectory starting at the optical surface of the Sun, it takes 1500 km 
to cross 100 g/cm$^2$ of matter. 
As a consequence, TeV pions and muons have plenty of time to decay 
before they lose energy independently
of the inclination of the shower trajectory: just the total depth 
is relevant to obtain the yield of secondary particles. As for the albedo flux of 
neutral particles escaping the Sun, it will depend on two main factors: how deep these particles 
are produced, and whether they are emitted outwards or inwards.

\begin{figure}[!t]
\begin{center}
\includegraphics[scale=0.59]{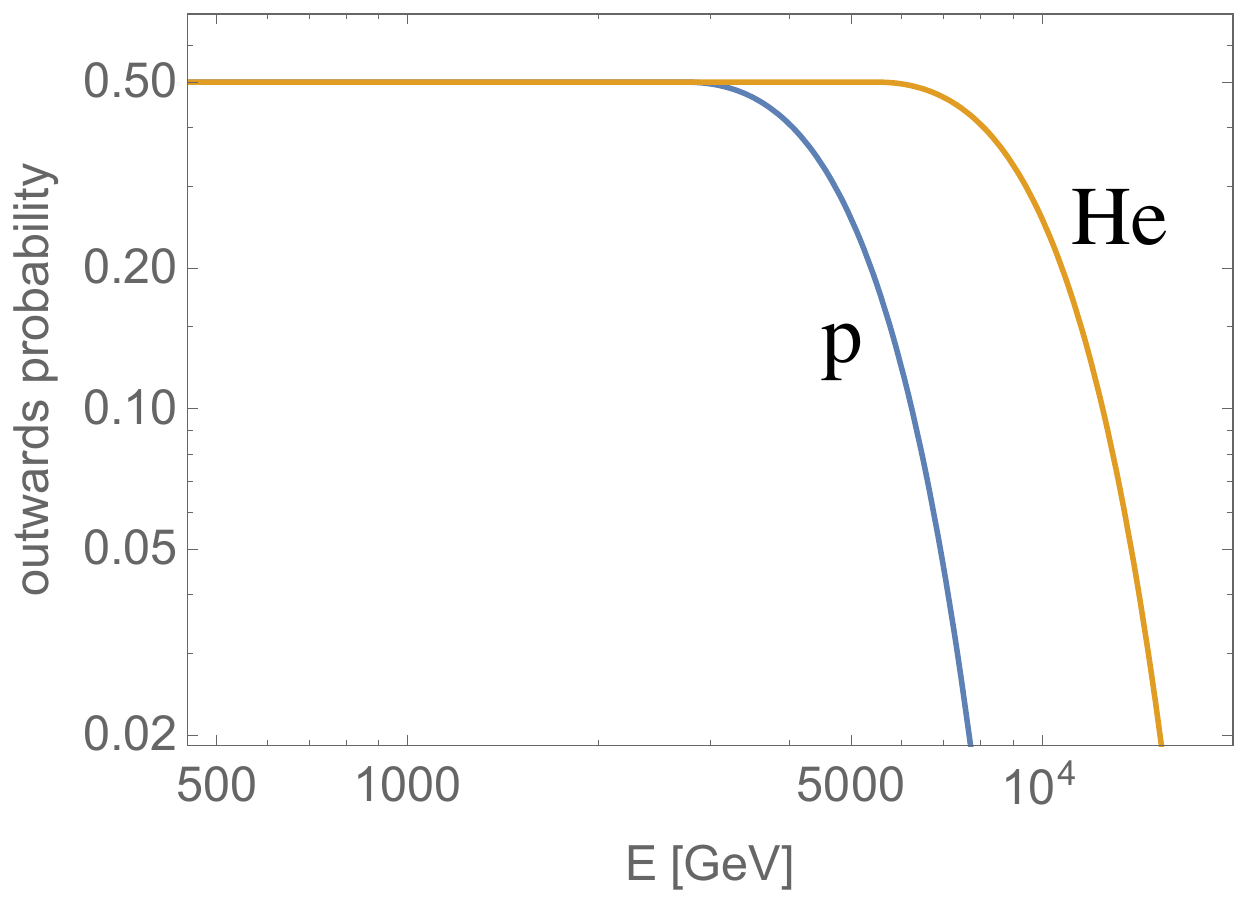}
\end{center}
\caption{Probability for an outwards emission by charge $\pm 1$ ($p,\;\pi$,...) 
and charge $+2$ (He) particles during an active phase of the Sun; for a 
quiet Sun we take $p^{\rm out}_{\rm quiet}(E)=p^{\rm out}_{\rm active}(3E)$.
\label{f2}}
\end{figure}

Our basic scheme is then the following. High energy CRs reach the solar 
surface aligned with an open (radial) field
line, and once there they start a shower. 
Secondary charged particles will continue penetrating the Sun as long as their 
energy is above some critical value $E_{\rm crit}\approx 10$ TeV that depends
on the phase in the solar cycle, whereas charged particles of lower energy 
will be trapped by closed
field lines and will shower at the approximated depth where they are produced.
Consistent with that, neutral particles 
produced in collisions and decays of charged particles 
will head outwards or inwards as a function of the energy of the
parent particle: above  $E_{\rm crit}$ 
most of the emission is inwards, whereas at lower energies the trajectory of 
charged particles is isotropized and so is their emission. 
In Fig.~\ref{f2} we plot the probability for an outwards emission during an active
phase of the Sun for different values of 
the parent energy. The neutral particles produced outwards, in turn, will 
propagate and eventually emerge from the Sun's surface. 

As the Sun goes into a quiet phase all magnetic effects will decrease. Qualitatively
this is what we observe at much lower  energies in the flux of galactic CRs, with
a larger fraction of 1--10 GeV CRs able to penetrate the heliosphere during those periods.
In terms of a critical energy analogous to the one we are using, a
reduction by a factor of 1/3 would imply an increase in the total CR energy by a factor 
of $3^{0.7}=2.1$, which seems consistent with the observations \cite{Zhu:2018jbk}.
Therefore, for a quiet Sun we will assume the same type of reduction by 
a factor of 1/3 in $E_{\rm crit}$; this will increase the CR flux absorbed
by the Sun (thin lines in Fig.~\ref{f1}) and will favour 
the inwards emission of neutral particles within the solar showers.

One important point about these fluxes concerns their distribution on the
Sun's surface seen from the Earth. In our scheme TeV
CRs approach the surface aligned with the radial field lines, which suggests spherical 
symmetry in the absorption ({\it i.e.}, same CR shadow seen from any direction). 
Moreover, most gammas and neutrons in this solar albedo flux
will be produced very close to
the surface (60\% of the absorbed CR energy is processed there), 
which implies that each point on the surface emits these particles
isotropically
(notice that gammas produced deeper will emerge preferentially along the radial direction).
An isotropic gamma emission from each point on the Sun's surface would imply that
at the Earth we see the peripheral regions brighter than the central ones, being the
distribution 
\beq
\Phi_\gamma(r) =\Phi_\gamma(0) {1\over \sqrt{1-{r^2\over R_\odot^2}}}\,
\eeq
with $\Phi_\gamma$ the gamma flux and $r$ the transverse distance from the
center.
However, during an active solar phase the open field lines tend to be pushed 
towards the 
polar regions, which would favor the CR absorption and the 
gamma emission from there. In any case, the current volume
of data seems insufficient to reach any clear conclusion on this point \cite{Linden:2018exo}.

\begin{figure}[!t]
\begin{minipage}{0.5\linewidth}
\begin{center}
\includegraphics[scale=0.5]{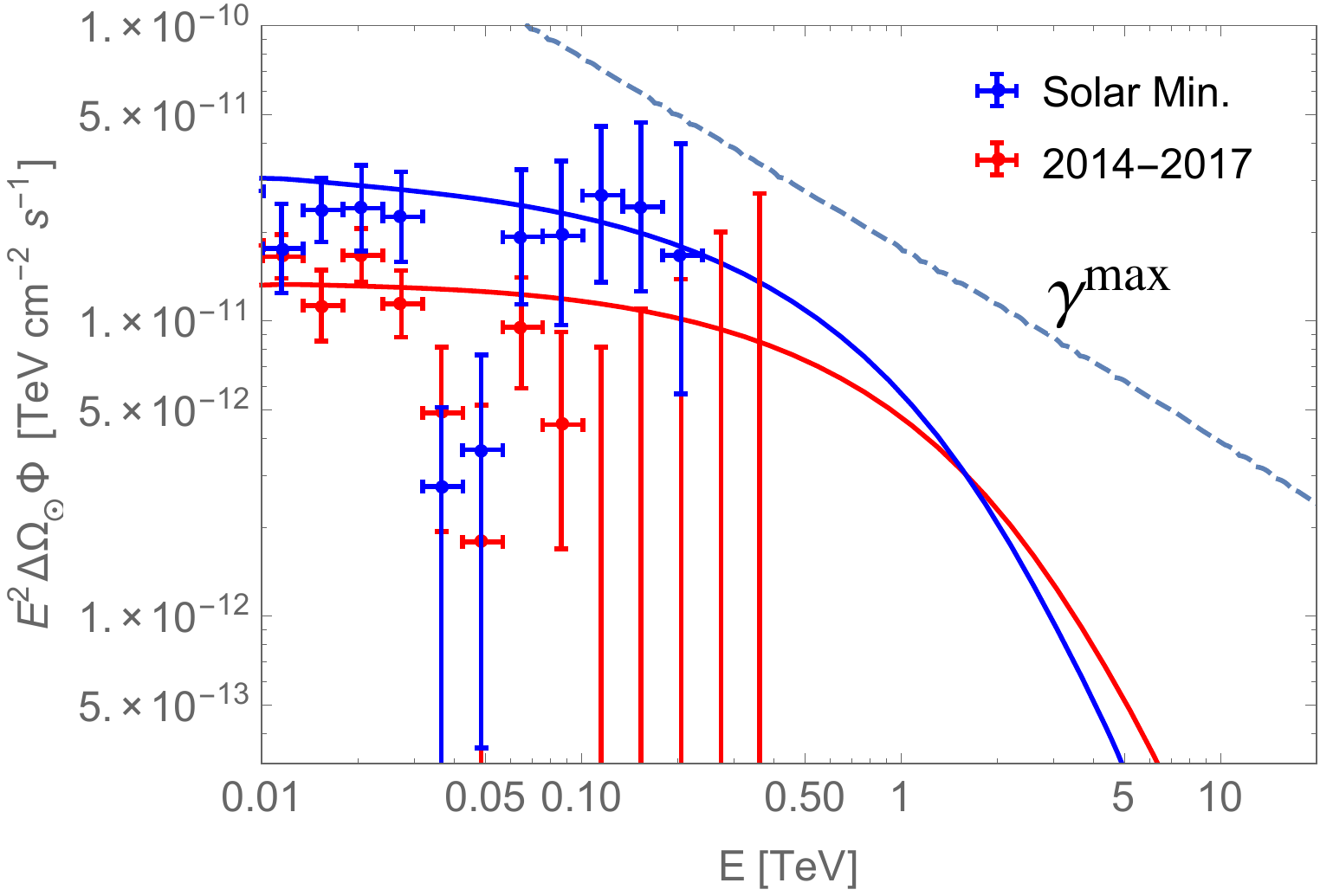}
\end{center}
\end{minipage}
\begin{minipage}{0.5\linewidth}
\begin{center}
\includegraphics[scale=0.5]{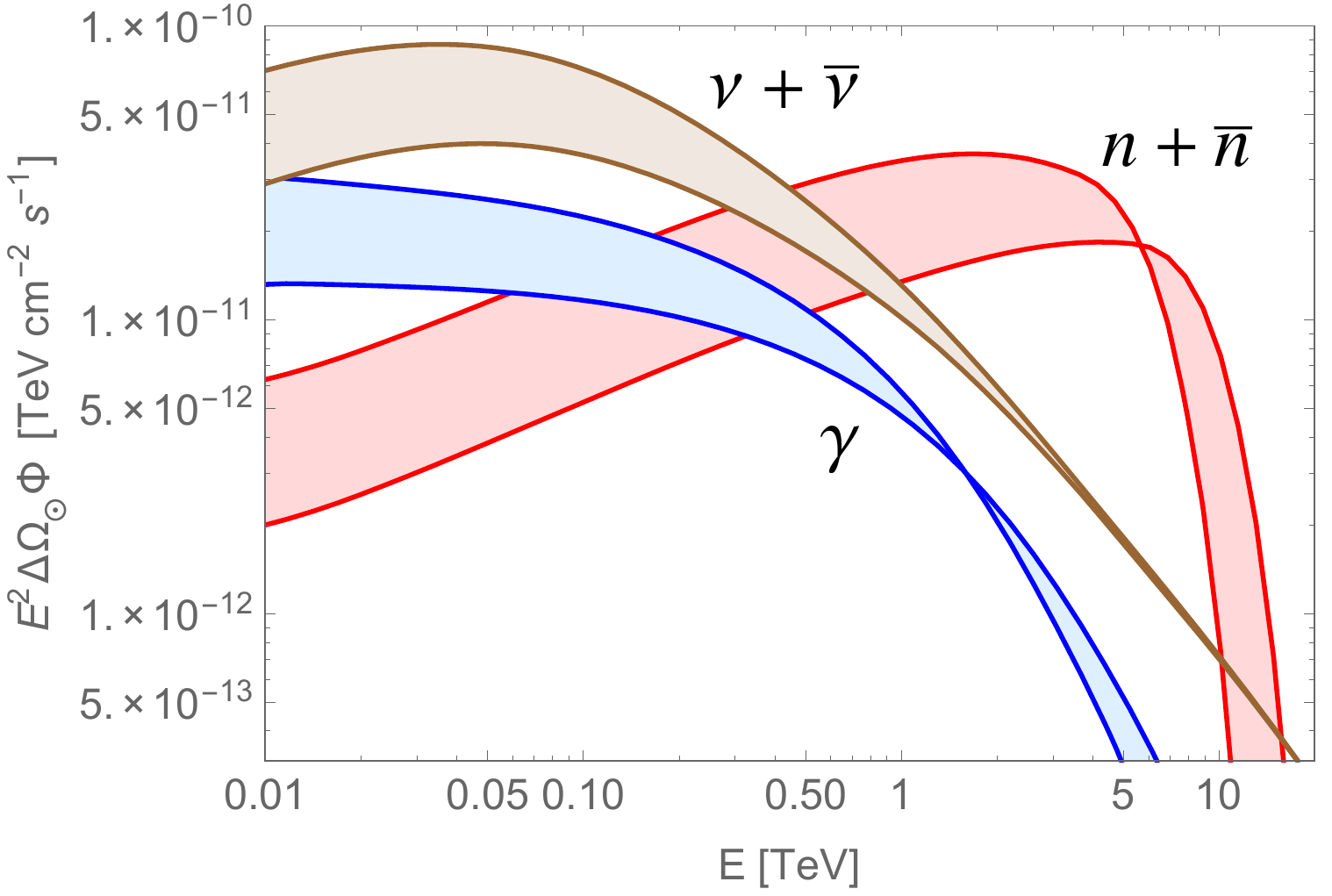}
\end{center}
\end{minipage}
\caption{{\bf Left.} Integrated flux of gammas that we estimate for the absorbed CR fluxes in 
Fig.~1 together with the Fermi-LAT data  (from \cite{Linden:2018exo}). The dashed line 
indicates the maximum gamma
flux (obtained if the CRs of all energies showered towards the Earth).
{\bf Right.} Fluxes of neutral particles observed at the Earth for the two CR fluxes in
Fig.~1 (including the gamma flux plotted on the left).
\label{f3}}
\end{figure}

To obtain the flux of neutral particles coming from the Sun we have parametrized
the yields in hadron collisions \cite{Pierog:2013ria}
and decays \cite{Illana:2010gh} at different energies and have solved cascade
equations \cite{Gaisser:1990vg}
for 17 species, including the electromagnetic component of the shower 
\cite{Gamez:2019dex}. 
For neutrinos, we have added the inwards flux 
produced at the hidden side of the Sun. Our results are summarized in Fig.~\ref{f3}. 
On the left we plot the integrated 
gamma flux that we obtain for $E>10$ GeV and the two 
absorbed fluxes in Fig.~\ref{f1} together with the two sets of 
Fermi-LAT data. At low
energies the gamma flux is reduced because most CRs are unable to reach
the Sun and shower there, while at $E>1$ TeV it drops because, although the parent
CRs always reach the surface, the gammas are produced towards the Sun and never reach
the Earth. Our estimate for a quiet Sun (blue line) relative to the active one (red line)
shows two different effects:
the gamma flux increases at low energies (CRs of lower energy can reach the surface
during that phase)
but it also drops faster at high energies (parent CRs 
have a larger tendency to emit the gammas inwards). Notice also that
our framework can not accommodate
a 40 GeV dip \cite{Tang:2018wqp} in the solar gamma flux.

On the right-hand side of Fig.~\ref{f3} we give the fluxes for the three
neutral species (gammas, neutrons and neutrinos) coming from the Sun. The neutrino flux is
basically twice the gamma one, as neutrinos of $E<1$ TeV produced in the hidden side of the Sun
can also emerge and reach the Earth. At higher energies the albedo neutrino
flux vanishes, but neutrinos from the peripheral regions in the hidden side of the Sun can still
emerge and reach us. At $E\approx 5$ TeV the neutrino flux becomes independent from
the solar cycle and has a spectral index around $-3.3$.
The total neutrino  flux that we obtain is above the
atmospheric background at $E>100$--$600$ GeV, depending on the zenith angle and
the solar phase;
at 5 TeV it is 20 times (4 times) larger than the atmospheric one from vertical
(horizontal) directions. Our results for the solar neutrinos are 
qualitatively similar to the ones obtained by SSG, and at $E>1$ TeV 
they agree within a $50\%$ with the results 
in  \cite{Edsjo:2017kjk,Ng:2017aur,Arguelles:2017eao,Masip:2017gvw}.
We estimate that
the Sun provides around 2 events per year in a km$^3$ telescope like 
IceCube \cite{Achterberg:2006md} or KM3NeT \cite{Adrian-Martinez:2016fdl}.

The solar neutron flux  has a very peculiar energy distribution: it mimics the shadow of
the Sun observed by HAWC up to 10 TeV and then it drops sharply. 
The origin of most of these neutrons is the spallation of He nuclei reaching the 
Sun's surface, being neutrons and antineutrons from hadronic collisions just
10\% of the flux. The signal consists of 100-300 neutrons/(year m$^2$) 
with a very hard spectrum (see Fig.~\ref{f3}). Our result is a factor of 20 larger
than the SSG estimate  \cite{Seckel:1991ffa}.
In a satellite based 
detector, a neutron event would provide a distinct signal with no 
track in the electromagnetic calorimeter but a TeV energy
deposition in the hadronic calorimeter.
The angular resolution would be poor but, since no cosmic flux of TeV neutrons
may be expected,
the background (albedo flux from the Earth's atmosphere) would be small.

The Sun is 
one of the brightest objects in the sky at TeV energies in the neutrino 
and the gamma channels. It is also the only possible source of TeV neutrons: since 
they have a $881$ s lifetime, at these energies 
they can not reach us from anywhere else. In addition, the Sun generates a 
shadow in the otherwise isotropic (at the 0.1\% level) CR flux.
Of course, the observation of any of these TeV solar fluxes in 
ground based detectors is very challenging: gamma ray telescopes operate only
at night (HAWC is one of the few the exceptions), neutrino telescopes face large atmospheric 
backgrounds, and the
detection of a $0.54^\circ$ diameter object requires in all the cases a very 
good angular resolution. Satellite based detectors avoid some of these problems, which
has allowed EGRET \cite{Orlando:2008uk} or 
Fermi-LAT to detect solar gammas and makes the neutron 
flux look accessible. 
Our work here correlates the signals in these four channels, in particular, we show 
that the use
of the CR shadow eliminates the main uncertainty (namely, 
the effects of the solar magnetic field) that plague previous
calculations. The discovery of the neutron and neutrino signals, plus more complete
results in the gamma and CR channels, would 
provide important information about the magnetic field,
the cycle and the interior of the Sun.

\section*{Acknowledgments}

This work has been supported by MICINN of Spain 
(FPA2016-78220, RED2018-102661-T, FPA2017-90566-REDC) 
and by Junta de Andaluc\'\i a (SOMM17/6104/UGR and FQM101).
MM would like to thank the Mainz Institute for Theoretical Physics 
of the Cluster of Excellence PRISMA+ (Project ID 39083149) 
and the Institute for Basic Science (Daejeon, Korea) 
for its hospitality and support.

\end{document}